# Unidirectional charge orders induced by oxygen vacancies on SrTiO$_3$(001)

Cui Ding,[1,2] Wenfeng Dong,[1] Xiaotong Jiao,[1] Zhiyu Zhang,[1] Guanming Gong,[1] Zhongxu Wei,[3] Lili Wang,[1,4] Jin-Feng Jia,[2,3,5] and Qi-Kun Xue[1,2,3,4]

[1]State Key Laboratory of Low-Dimensional Quantum Physics, Department of Physics, Tsinghua University, Beijing 100084, China

[2]Quantum Science Center of Guangdong-HongKong-Macao Greater Bay Area, Shenzhen 518045, China

[3]Department of Physics, Southern University of Science and Technology, Shenzhen 518055, China

[4]Frontier Science Center for Quantum Information, Beijing 100084, China

[5]Department of Physics and Astronomy, Shanghai Jiao Tong University, Shanghai 200240, China

[#]Corresponding authors: liliwang@tsinghua.edu.cn, jfjia@sjtu.edu.cn, qkxue@tsinghua.edu.cn

**Abstract**

The discovery of high-mobility two-dimensional electron gas and low carrier density superconductivity in multiple SrTiO$_3$-based heterostructures has stimulated intense interest in the surface properties of SrTiO$_3$. The recent discovery of high-$T_c$ superconductivity in the monolayer FeSe/SrTiO$_3$ aroused the upsurge and underscored the atomic precision probe of the surface structure. By performing atomically resolved cryogenic scanning tunneling microscopy/spectroscopy characterization on dual-TiO$_{2-\delta}$-terminated SrTiO$_3$(001) surfaces with ($\sqrt{13} \times \sqrt{13}$), c(4 × 2), mixed (2 × 1), and (2 × 2) reconstructions, we disclosed universally broken rotational symmetry and contrasting bias- and temperature-dependent electronic states for apical and equatorial oxygen sites. With the sequentially evolved surface reconstructions and simultaneously increasing equatorial oxygen vacancies, the surface anisotropy reduces, and the work function lowers. Intriguingly, unidirectional stripe orders appear on the c(4 × 2) surface, whereas local (4 × 4) order emerges and eventually forms long-range unidirectional c(4 × 4) charge order on the (2 × 2) surface. This work reveals robust unidirectional charge orders induced by oxygen vacancies due to strong and delicate electronic-lattice interaction under broken rotational symmetry, providing insights into understanding the complex behaviors in perovskite oxide-based heterostructures.

Keywords: SrTiO$_3$(001), surface reconstruction, oxygen vacancy, unidirectional charge order,

scanning tunneling microscopy/spectroscopy, work function

Pristine SrTiO$_3$ is an insulating cubic perovskite, with a = 3.905 Å at room temperature in its bulk form.[1] Below 105 K, the antiferrodistortive transition, due to the antiphase rotation of adjacent TiO$_6$ octahedra around one cubic axis, induces a weakly tetragonal structure, and correspondingly, in-plane twofold symmetry.[1, 2] Independent of bulk carrier density (either semiconductivity in pristine bulk or metalicity in doped bulk), SrTiO$_3$ surfaces and SrTiO$_3$-based heterostructures host high-mobility two-dimension (2D) electron gas ($10^{13}$-$10^{14}$cm$^{-2}$)[3, 4] and ultra-low carrier density ($10^{13}$ cm$^{-2}$) 2D superconductivity.[5-7] Despite previous intensive experimental investigations using various state-of-the-art techniques,[4, 8-20] the origin of the surface/interface conductance remains an ongoing debate. Modulation doping with oxygen vacancies as the intrinsic donors[4, 8, 13-15, 21] and polarization doping due to oxygen vacancies tuned distortion[16-18] had been hotly debated, and large polarons had been proposed to be key quasiparticles in SrTiO$_3$-based surfaces/interfaces.[19] Moreover, the surface/interface conductivity shows even 1D features,[20] in addition to 2D nature with high tunability under various stimuli (such as strain, electric field, photon illumination, and step/defect engineering),[22] revealing close interactions between multiple degrees of freedom including lattice, spin, charge, orbital and polar.

Structurally, TiO$_2$-termination is particularly desirable for conductive interfaces/surfaces, which is favorable under the reduction process, i.e., ultra-high vacuum annealing.[10, 13, 15] Accompanied by surface oxygen desorption under annealing at temperatures above 600°C, SrTiO$_3$(001) surfaces exhibit dual-TiO$_{2-\delta}$ termination, wherein corner- and edge-sharing truncated TiO$_{5\square}$ ($\square$ a vacant oxygen site, i.e., oxygen vacancies) octahedral units mesh atop a bulk-like TiO$_2$ base-layer composed of TiO$_6$ octahedra.[23-25] As illustrated in Fig. 1(a), TiO$_{5\square}$ octahedra could arrange into recurrent structural motifs of Z-pentamer, giving rise to ($\sqrt{13} \times \sqrt{13}$)-R33.7° reconstruction.[25-27] Upon TiO$_{5\square}$ concentration increased to same as the TiO$_6$ octahedra in the base layer, c(4 × 2) (Fig. 3(a)), (2 × 1) and two-types (2 × 2) (Figs. 3(b,c)) reconstructions, composed of compact TiO$_{5\square}$ tetramer (Figs. 3(a, c)), dimer and trimer (left and right half-panels in Fig. 3(b), respectively), form instead.[24, 25] The equatorial oxygen sites at the edges (O$_e$, above the hollow sites of base-layer TiO$_6$) and the corners (O$_c$, right above base-layer TiO$_6$) of various adlayer TiO$_{5\square}$ motifs have relatively low bond valence sum and, therefore, readily desorb and leave oxygen vacancies.[25] On the other hand, the exposed apical oxygen sites of the base-layer TiO$_6$ octahedra (O$_a$) could introduce varied Ti-$d$ states as well.[26, 27]

Here we conducted controlled ultra-high vacuum annealing to obtain (√13 × √13)-33.7°, c(4 × 2), (2 × 1), and (2 × 2) reconstructed SrTiO$_3$(001) surfaces and then performed cryogenic scanning tunneling microscopy/spectroscopy (STM/S) characterization to achieve atomic-scale probes of surface structures. We identified unidirectional charge orders induced by oxygen vacancies under broken rotational symmetry. Our results reveal strong and delicate atomic-site-dependent electronic states and electronic-lattice interaction that gives fruitful nanoscale features as well as phase separation, shedding light on understanding the complex behaviors in perovskite oxide.

**Results**

**Twofold symmetry with chirality and bias-dependent modulation on (√13 × √13) surface**

We start from the resolved (√13 × √13) surface to acquire the primary information on the dual-TiO$_{2-\delta}$ surface, for the similar surface states at LN and LHe temperatures,[26, 27] in striking contrast to the varied electronic states on other reconstructions, to be discussed later. Here, we presented the results at LHe temperature. In each (√13 × √13) unit, ten edge-shared TiO$_{5\square}$ octahedra in the adlayer stack on twelve TiO$_6$ plus one center TiO$_{5\square}$ octahedra in the base-layer, with reduced bonds between the adlayer Ti to the base-layer oxygen compared with bulk.[28] Considering the low-temperature tetragonal structure with a random c axis, we define the surface [100] direction as the one of a small surface lattice (i.e., $a_0 < b_0$) and in-plane anisotropy ratio γ = $b_0/a_0$. Correspondingly, as depicted in Fig. 1(a), the A border of the (√13 × √13) units, spanning over the Z-pentamer marked in blue with [100] directional major axis, is shorter than the B border marked in purple. Figure 1(b) displays the atomically resolved image taken around a twin boundary at a sample bias $V_s$ = 1.0 V, showing checkerboard patterns consisting of dark and faint dark tiles separated by the Z-pentamer frame. The inserted fast Fourier transformation (FFT) images show ($q_A$, $q_B$) Bragg spots and (√2$q_A$/2, √2$q_B$/2) modulation peaks with consistent $q_A > q_B$, where ($q_A$, $q_B$) is the reciprocal vector of (√13 × √13). Associated with reversed lattice anisotropy in the adjacent domains around the twin boundary, the (√13 × √13) units exhibit mirror symmetry with reversed chirality of TiO$_{5\square}$-pentamer (Supplementary Materials Fig.S1), as illustrated in Fig. 1(a).

Due to the orbital selectivity of Ti-$d$ states,[26, 27] the adlayer TiO$_{5\square}$ Z-pentamers dominate the density of states (DOS) under $V_s$ < 2.0 V (Fig. 1(c) at $V_s$ = 1.0 V and Fig. 1(d) at $V_s$ = 1.5 eV), whereas the base-layer TiO$_{5\square}$ contribute dominant DOS at larger bias (Fig. 1(e) at $V_s$ = 4.0 V, and more data in Supplementary Materials Fig. S1). Notably, (√2$q_A$/2, √2$q_B$/2) modulation peaks are synchronized with the high DOS state of Z-pentamers, i.e., strong under $V_s$ < 2.0 V but almost invisible upon reversed high-DOS under larger bias (Supplementary Materials Fig. S1). The scattered small dots on borders correspond to equatorial oxygen vacancies at the inner corner sites of the Z-pentamer (O$_e$ in Fig. 1(a)), which preferentially occur on relatively short borders and induce local elongation of 1 Å, as exemplified in Fig. 1(c). Compared with the images in Figs. 1(b-e), the detailed DOS contrasts of the Z-pentamers are resolved under a special tip state (Fig. 1 (f)), agreeing well with previous simulation results.[27] As resolved from the work function mapping image displayed in Fig. 1(g), the adlayer Z-pentamer induces lowered work function values ($\varphi$), and O$_e$ vacancy leads to further lowering with a spatial shift of ~ $a_0$ (the local minimum marked by the white dashed line relative to the oxygen vacancy marked by the red circle).

**Unidirectional alignment of TiO$_{5\square}$ motifs**

Upon annealing at elevated temperatures above 1000 °C, the surface sequentially evolves into c(4 × 2), (2 × 1), and (2 × 2) dominant reconstructions with different intermixing. The large-scale images in Figs. 2(a-d) show atomically flat terraces separated by steps of unit-cell high (3.90 Å) with characteristic edge orientations. The atomically resolved images taken on the terraces are displayed in Figs. 2(e-h), with the inserted fast Fourier transformation (FFT) images showing the c(4 × 2), (2 × 1), and (2 × 2) Bragg spots, respectively, and the corresponding inverted-FFT images in Figs. 2(i-l). The zoom-in atomically resolved images in Figs. 2(m-p), with atomic correspondence in representative unit cells marked, show consistent high DOS of the adlayer TiO$_{5\square}$ motifs at $V_s$ = 0.5 V. The surface reconstructions vary in nanometer scale (Figs. 2(i-l)), resulting in twin (anti-phase) boundaries with inversed anisotropy (staggered units) between the adjacent domains, as exemplified by white (yellow) dashed lines in Figs. 2(f,j) (More results in Supplementary Materials Fig. S2).

The reconstructions share the common twofold symmetry with large periodicity along the relatively small lattice direction, as evidenced by $4a_0 \times 2b_0$ for two sets of c(4 × 2) (Fig. 2(e)) and $2a_0 \times b_0$ for the two sets of (2 × 1) units (Fig. 2(f)) under the definition of $b_0 > a_0$. Given the broken rotational symmetry of the patterns illustrated in Figs. 3(a,b), the TiO$_{5\square}$ motifs preferentially align densely along the large-lattice direction [010], which is reasonable regarding the negatively charged nature of the truncated TiO$_{5\square}$ octahedra. On the (2 × 2) surface, additional c(4 × 4) order occurs and exhibits unidirectional features as well, evidenced by the unidirectional dimers (exemplified by the dots-lines in Fig. 2(l)), the details to be discussed later. The preferential large-lattice-directional feature, together with varying surface domains, leads to the characteristic steps displayed in Figs. 2(a-d), i.e., [010] directional step with minority [100] fragments for the c(4 × 2) surface, [110] directional step consisting of alternative fragmental nanometer-long [010] and [100] edges for the (2 × 1) surface, and [110] directional step with sprinkled kinks for the (2 × 2) surface.

**Temperature- and bias- dependence on c(4 × 2) and plain (2 × 2) surfaces**

In striking contrast to the relatively localized states at LN temperature (Figs. 2 (e-h, m-p), the surface electronic states evolve to delocalize with significantly increased DOS upon lowered temperature of 4.8 K, introducing various electronic orders with significant bias-dependence. Figures 3(d-g) and Figs. 3(j-m) display the bias-dependent atomically resolved images for the c(4 × 2) and (2 × 2) dominant surfaces, respectively, while Figs. 3(i) and 3(o) the respective work function mapping images of the same regions in Figs. 3(h) and 3(n). Consistent with the bias-dependent contrast observed on the (√13 × √13) reconstruction (Figs. 1(c-e)), some O$_a$ sites (marked by yellow dashed circles) show bright contrast at $V_s$ = 4.0 V but faint contrast at 0.5 V, whereas the adlayer TiO$_{5\square}$ motifs contribute strong DOS at $V_s$ = 0.5 V instead, with particularly high DOS contrast around some sites marked by red and purple dashed circles. The high DOS around the Fermi level corresponds to lowered work function (Figs. 3(i, o)). Those marked by purple dashed circles introduce much higher DOS and lower $\varphi$ than those marked in red, which will be disclosed to correspond to paired O$_e$ vacancies and single O$_e$ vacancies, respectively. The spatial displacements in local work function minimums (as exemplified in Fig. 3(i), marked by white dashed lines) are consistent with $O_e$ vacancies on (√13 × √13) (Fig. 1(g)).

On the c(4 × 2) surface, the scattered $O_c$ vacancies (black dashed circles), featured by high DOS at 4.0 V and lowered but still bright contrast at 0.5 V, preferential align along the [100] direction with an unidirectional interval of at least $2a_0$ and mostly $4a_0$. Under the small bias of 0.5 V, the c(4 × 2) units could exhibit unidirectional stripes along one diagonal direction accompanied by $O_e$ vacancies, as exemplified by gray dashed lines in Figs. 3(g, h) (More results in Supplementary Materials Fig. S3). On the (2 × 2) dominant surfaces, those large-bias high-DOS $O_a$ sites align along each lattice direction with the nearest distance of $4a_0$, giving (4 × 4) or partial c(4 × 4) periodicity under in-phase and out-of-phase alignments in the every other lattice rows, respectively, as exemplified by light blue dashed squares and diamonds in Figs. 3(j, k). As a result of the locally reversed lattice anisotropies, the inserted FFT patterns exhibit two sets of ($q_a/4$, $q_b/4$) peaks. At $V_s$ = 0.5 V, the (2 × 2) units are resolved as quatrefoils, marked by colored dashed squares with dashed circles at the centered $O_a$ sites, with central symmetry (Figs. 3(l-n)), supporting the type-II (2 × 2) model illustrated in Fig. 3(c). That is, the (2 × 2) adlayer consists of the same corner-shared $TiO_{5\square}$ tetramers along the [010] direction of the large lattice as the c(4 × 2) adlayer, but with in-phase, instead of out-of-phase, alignment along the [100] direction of the small lattice.

Combining the above bias-dependent atomically resolved topographic images and the work function mapping image in Fig. 3(o), the (2 × 2) units are classified into three groups featured by distinct electronic features. Those characterized as bright (opposite) quatrefoils at 0.5 V (yellow (green) dashed squares), with high (faint)-contrast $O_a$ sites at 4.0 V (yellow (green) dashed circles), exhibit medium (lowest) work function values. Those marked in green are usually accompanied by $O_e$ vacancies. The two kinds (2 × 2) units alternately arrange, resulting in the (4 × 4) or partial c(4 × 4) orders resolved at large bias, as exemplified in Figs. 3(k, m). Those at the center of the (4 × 4) periodicity with the faintest $O_a$ sites marked in orange exhibit the highest work function values. Further zoom-in images showing the three types (2 × 2) units are displayed in Supplementary Materials Fig. S4. Notably, quasi-quatrefoils with contrasting

DOS spatial extensions appear in Figs. 3(f, g), likely corresponding to dispersed (2 × 2) units.

**Unidirectional c(4 × 4) charge order on (2 × 2) surface**

The most important finding is the c(4 × 4) charge order emerged on the (2 × 2) surface upon further oxygen desorption under elongated moderate annealing. At $V_s$ = 0.5 V, a precursor c(4 × 4)-I (marked by navy blue diamonds in Fig. 4(b)), composed of alternately arranged (2 × 2) units featured as bright and faint quatrefoils, occurs around the aforementioned (4 × 4) order (light blue squares) on the plain (2 × 2) surface. Whereas, the universal c(4 × 4) order over ten nanometers displayed in Fig. 2(h) and Fig. 4(c) consists of c(4 × 4)-II labeled in purple, originating from every second varied jointed corner shared by two $TiO_{5\square}$ tetramers along the large lattice direction that runs in out-of-phase along the small lattice direction, as illustrated by the purple dashed diamond in Fig. 4(a) with a relative shift of $a_0$ to c(4 × 4)-I. The combined atomically resolved topographic image and the work function mapping image in Figs. 4(g, h) disclose paired $O_e$ vacancies (paired purple dashed circles) around each corner of the c(4 × 4) unit cells, with similar features of $O_e$ vacancies. Notably, the adjacent shared $TiO_{5\square}$ tetramers are distinguishable at LN temperature (Figs. 2(h, l, p)), whereas paired $O_e$ vacancies exhibit spatially extended high-DOS at LHe temperature due to the increased delocalized DOS (Fig. 4(d)). At the large bias of 3.5 V, the $O_a$ sites displace the $O_e$ vacancy pairs to contribute to high-DOS, resulting in a structure correspondence same as c(4 × 4)-I (Figs. 4(a, e)). These results demonstrate that an alternative arrangement of (2 × 2) units with contrasting DOS, enclosed by the green dashed square (probably containing paired $O_e$ vacancies) and the yellow one in Fig. 4(a), gives the c(4 × 4) order.

**Interdependence of surface anisotropy, oxygen vacancy, and work function**

Statistically, the surface anisotropy ratio $\gamma = b_0/a_0$ is up to 1.07-1.19 for c(4 × 2) reconstruction, reduces to 1.03 for the plain (2 × 2) reconstruction, and further to 1.01 (1.03 at LN temperature) with the universal c(4 × 4) order (Supplementary Materials Table S1). Meanwhile, oxygen vacancies gradually increase (Supplementary Materials Table S2), correspondingly, the work function values decrease sequentially with an interval difference value of ~ 0.2 eV (Supplementary Materials Fig. S5). Assuming each $O_e$ vacancy denotes two electrons, the surface carrier density is estimated at the magnitude of $10^{13}$ cm$^{-2}$ for (√13 × √13) and c(4 × 2) surfaces and over $10^{14}$ cm$^{-2}$ for the plain (2 × 2) and c(4 × 4) surfaces (Supplementary Materials Table S2), agreeing well with previous photoemission results.[13, 15] In particular on the plain (2 × 2) and c(4 × 4) surfaces, despite similar reciprocal vectors from the FFT patterns of work function and topographic images (Supplementary Materials Fig. S5), the former exhibits unidirectional intensity contrast, i.e., the stronger intensity of the Bragg spots along the large lattice direction $I(q_a/2) > I(q_b/2)$ (Fig. 3(o) and Fig. 4(h)) and synchronized unidirectional intensity contrast in the c(4 × 4) order peaks (Fig. 4(h)). This fact confirms the robust unidirectional electronic states over the wide range of surface oxygen vacancies.

**Discussion and Conclusion**

Our atomically resolved cryogenic STM/STS characterization on dual-TiO$_2$-terminated SrTiO$_3$(001) surfaces, with (√13 × √13), c(4 × 2), (2 × 1), and (2 × 2) reconstructions, disclose broken rotational symmetry and unidirectional electronic states. The contrasting apical and equatorial oxygen vacancies lead to various bias-dependent surface electronic structures, and the latter contributes significantly increased DOS, with localized to itinerant features, from LN temperature lowered to LHe temperature. Moreover, the equatorial oxygen vacancy reduces adlayer anisotropy and surface work function locally to universally with increased intensity. These results demonstrate the strong electron-lattice coupling tuned by oxygen vacancies, which has been termed polarons,[15, 18, 29] and the delicate interplay with electronic interaction, which could result in varied electron- and spin-related novel properties.[30-34]

More importantly, the equatorial oxygen vacancies induce stripe order on the c(4×2) surface and (4 × 4)/c(4 × 4) orders on the (2 × 2) surface with robust unidirectional features, probably connected to electronic nematicity, a ubiquitous feature of iron-based superconductors and cuprates.[35, 36] Furthermore, the electronic 4a$_0$ order shares resemblances with the checkerboard charge orders in cuprates and monolayer FeSe that serve as local pairing species.[37-41] Our results indicate that the 4a$_0$ unidirectional charge order emerges upon self-oriented surface oxygen vacancies under strong electron-phonon coupling with rotational symmetry breaking, which populates in real space and eventually induces global coherence.

It is worth noting that the (2 × 2) surface exhibits Ti-$t_{2g}$ derived quasiparticle state (2D electron gas) located in an energy range of less than 300 meV below the $E_F$ with 2D carrier density spanning from $10^{13}$cm$^{-2}$ up to $10^{14}$cm$^{-2}$, accompanied by varied electron-phonon interaction from strong long-range in polaron regime (feature by photoemission replica bands) to weak short-range in conventional metallic states (feature by photoemission kinks, similar as in the case of high-$T_c$ superconducting cuprates).[13, 15] On the other hand, superconductivity emerges in the TiO$_2$ layer of SrTiO$_3$ under various stimuli (doped, reduced, gated, strained, and interfacial and polar engineered),[5, 22, 42-46] under a similar nominal 2D carrier density of $10^{13}$cm$^{-2}$, with distinct oxygen isotope effect and extensive tunability by oxygen vacancies.[47-49] Regarding the controllable electronic and electron-phonon coupling on the various reconstructions disclosed above, the dual TiO$_{2-\delta}$ terminated STO(001) surface is an ideal system to directly probe the pairing states under low carrier density (an unconventional feature of high-temperature superconductivity) and its evolution over a wide range of carrier density and electron-phonon coupling strength.

In summary, we disclose oxygen vacancies tuned surface distortion and electronic states from atomic-site dependence to universal order with increased density and self-orientation upon moderate reduction. The delicate interplay between electron interaction and electron-lattice coupling, the robust unidirectional electronic orders, and the ubiquitous nanoscale spatial phase separation presented in this work provide insights into understanding the complex behaviors in perovskite oxide, cuprates, and interface high-temperature superconductivity.

**Materials and Methods**

Our experiments were carried out in a Createc ultrahigh vacuum low-temperature STM system equipped with a molecular beam epitaxy chamber. The base pressure is better than $1.0 \times 10^{-10}$ Torr. The Nb-doped SrTiO$_3$(001) (0.05 wt. %) substrates were annealed at 850 °C, 1000 °C, and 1100 °C for 20 - 30 minutes to obtain atomically flat TiO$_{2-\delta}$-terminated surface with (√13 × √13), c(4 × 2), mixed (2 × 1) and (2 × 2) reconstructions, respectively. Upon elongating annealing above 600 °C, universal c(4 × 4) order occurs on the (2 × 2) reconstructed surface. All STM measurements were performed in a constant current mode (tunneling current set point $I_t$ = 100 pA) under liquid Helium (LHe, at 4.8 K) and liquid Nitrogen (LN, at 78 K), with a polycrystalline PtIr tip and the bias voltage applied to the sample. The sample bias ($V_s$) is labeled in each image. To get further insight into the surface electronic structure, we collected work function (d$\ln I$/d$z$) mapping images instead of d$I$/d$V$ tunneling spectra, for the latter show large semiconducting gaps spanning over the Fermi level (Supplementary Materials Fig. S7). The values of ln$I$ are measured while disabling the feedback circuit and decreasing the tip-sample distance $z$ under $V_s$ = 1.0 V, and then the d$\ln I$/d$z$ is extracted from the linear fitting of the ln$I$-$z$ relation. The local tunneling barrier height is calculated based on the equation $\phi = 0.95 \left(\frac{d\ln I}{dz}\right)^2$, wherein $I$ in the units of A and $z$ in the units of Å.

## Acknowledgments

This work is supported by the National Natural Science Foundation of China (Grants No. 52388201, 12074210 and 12204222), the National Key Research and Development Program of China (Grant No. 2022YFA1403102), and the Basic and Applied Basic Research Major Programme of Guangdong Province, China (Grant No. 2021B0301030003) and Jihua Laboratory (Project No. X210141TL210).

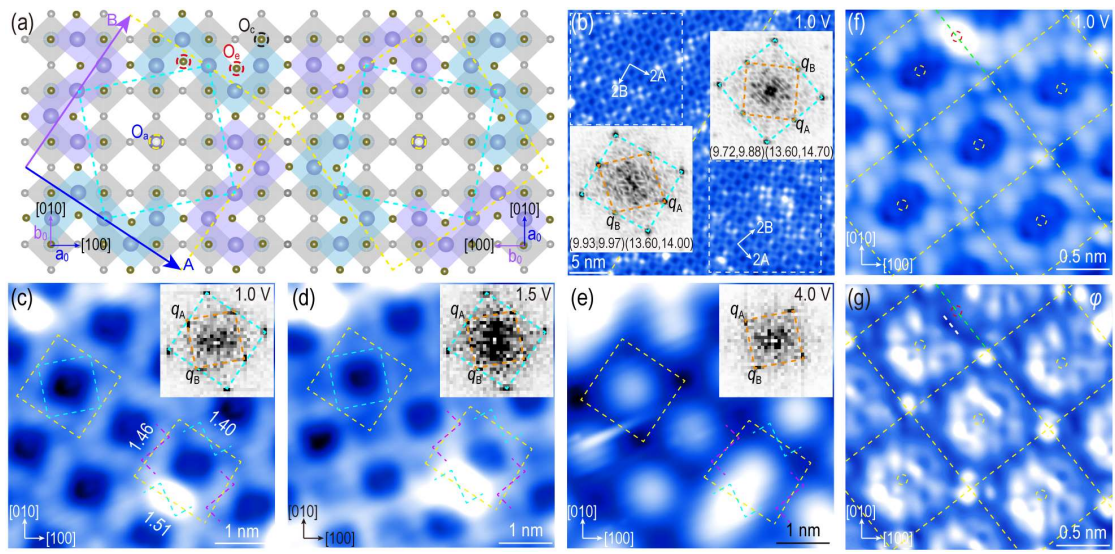

Fig.1 LHe-temperature characterization of SrTiO$_3$(001)-(√13 × √13) surface. (a) Schematic illustration of mirrored SrTiO$_3$(001)-(√13 × √13) reconstructions consisting of chirality reversal TiO$_{5\square}$ Z-pentamer. (b) Atomically resolved topographic image around a twin boundary marked by the yellow dashed line, with the respective FFT images corresponding to the regions marked by the white dashed squares in the adjacent domains inserted. The numbers are the deduced ($2\pi/q_A$, $2\pi/q_B$) and ($\sqrt{2}\pi/q_A$, $\sqrt{2}\pi/q_B$) values in the units of Å. (c, d, e) Bias-dependent atomically resolved topographic images. (f) Atomically resolved topographic image and (g) work function mapping image of the same region. The colored dashed diamonds/circles mark the respective units/oxygen vacancies illustrated in (a), and the white dashed line in (g) marks the nearby work function minimum around an O$_e$ vacancy.

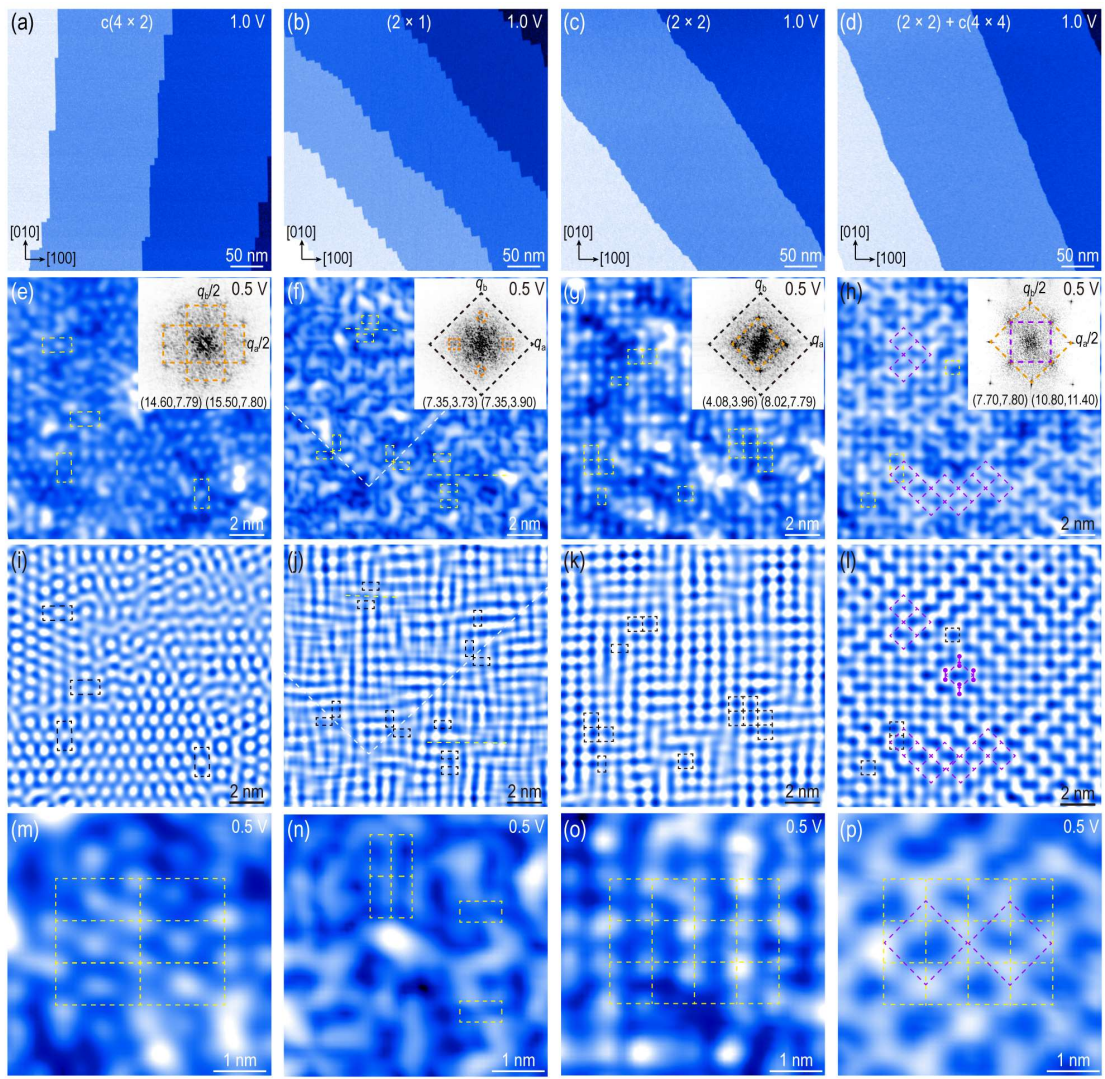

Fig. 2 LN-temperature characterization of SrTiO$_3$(001)- c(4 × 2), -(2 × 1), -(2 × 2) and -(2 × 2) + c(4 × 4) surfaces. (a, b, c, d) Large-scale topographic images showing the terrace-step structures. (e, f, g, h) Atomically resolved topographic images taken on terraces with the respective FFT images inserted, and (i, j, k, l) the corresponding inverted FFT images generated from all the Bragg spots, respectively. The numbers at the bottom of FFT images are the deduced ($2n\pi/q_a$, $2m\pi/q_b$) values in the units of Å. (m,n,o,p) Zoom-in atomically resolved topographic images. The yellow/black dashed squares/rectangles mark the respective unit cells, the white (yellow) dashed lines the twin (anti-phase) boundaries, and the purple dashed diamonds the c(4 × 4) units.

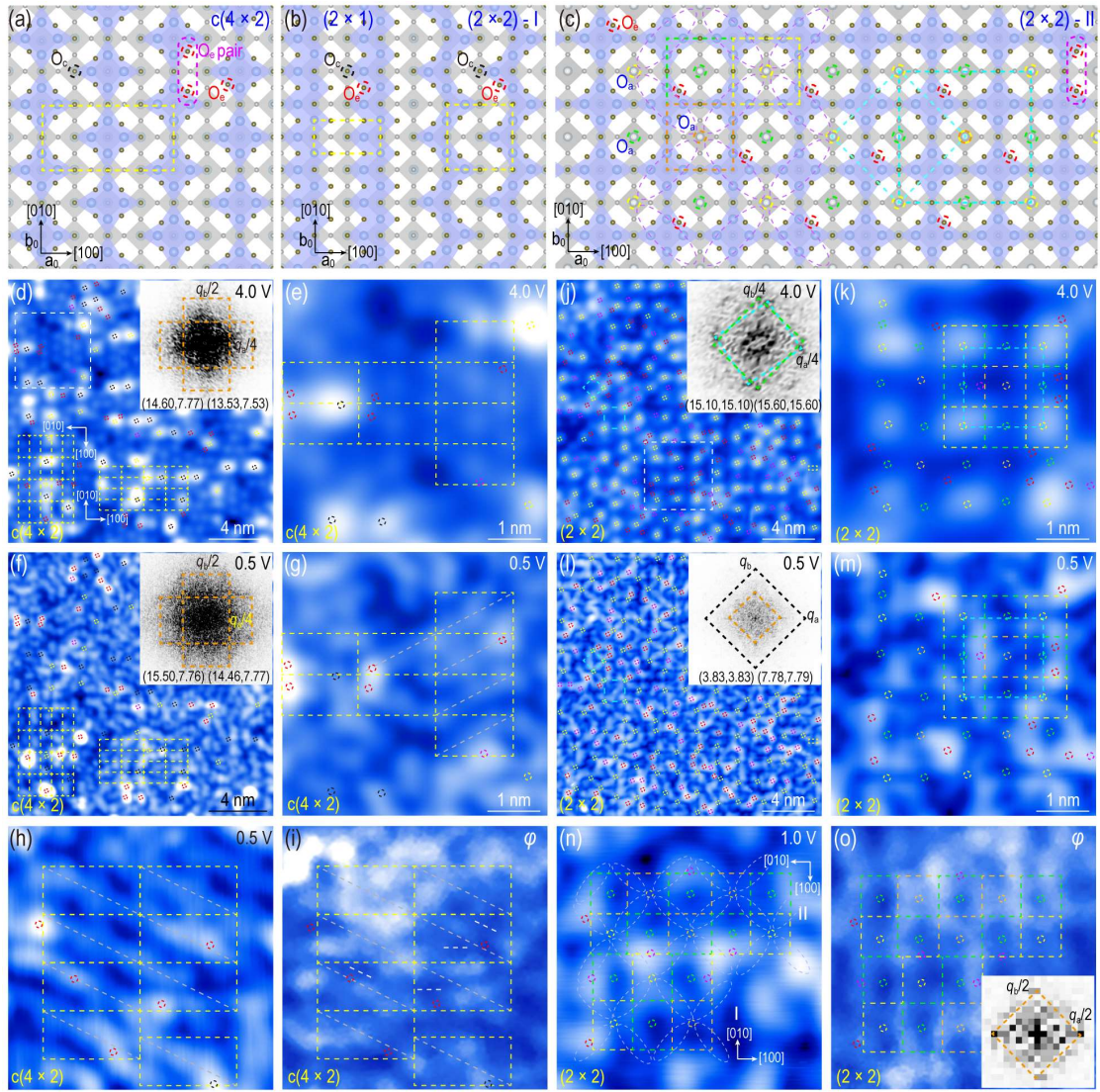

Fig.3 LHe-temperature STM characterization of SrTiO$_3$(001)- c(4 × 2) and plain (2 × 2) surfaces. (a,b,c) Schematic illustration of c(4 × 2), (2 × 1) and (2 × 2)-I, and (2 × 2)-II reconstructions, respectively. (d-g) and (j-m) Bias-dependent atomically resolved topographic images of c(4 × 2) and (2 × 2) dominant surfaces, respectively; (d,f) and (j,l) the large-scale images with corresponding FFT images inserted, and (e,g) and (k,m) the zoom-in images of the regions marked by respective white dashed squares in (d) and (j). The colored dashed rectangles/squares enclose the periodic units, the purple circles/ovals mark the paired O$_e$ vacancies, the black/red circles single O$_c$/O$_e$ vacancies, and the other circles single specific apical oxygen sites, as depicted in (a-c). (h) Atomically resolved topographic image and (i) work function mapping image of the same region on c(4 × 2) surface, while (n) and (o) on (2 × 2) surface. The white dashed lines in (i) mark the local work function minimums around O$_e$ vacancies.

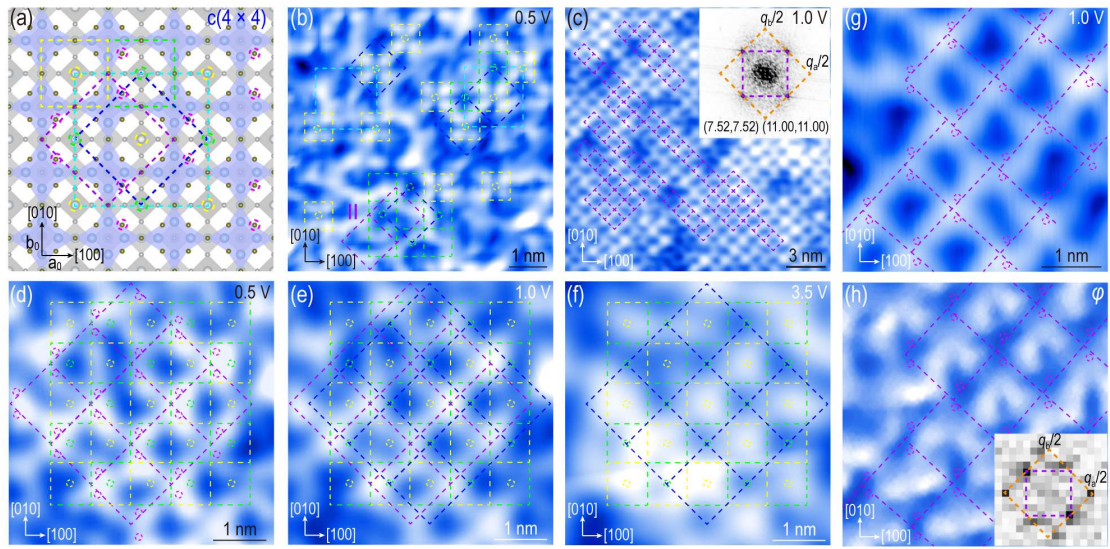

Fig. 4 LHe-temperature characterization of c(4 × 4) charge order on SrTiO$_3$(001)-(2 × 2) surface. (a) Schematic illustration of c(4 × 4) on (2 × 2). (b) and (c) Atomically resolved topographic images showing the two types c(4 × 4) orders at the primary state and long-range c(4 × 4)-II order formed eventually. (d, e, f) Bias-dependent atomically resolved topographic images showing the DOS displacement. (g) Atomically resolved topographic image and (h) work function mapping image of the same region with the corresponding FFT pattern inserted. The colored dashed squares/diamonds mark the (2 × 2)/(4 × 4)/c(4 × 4) units, the purple dashed circles mark the single O$_e$ vacancies, and the yellow/green circles mark the apical oxygen sites of the respective (2 × 2) units in the same color, as depicted in (a).

**Supporting information**

Statistics of surface anisotropy ratios, densities of oxygen vacancies, and work function values of various reconstructions, in addition to more bias-dependent atomic resolution images.

**References**

(1) Cowley, R. A. Lattice Dynamics and Phase Transitions of Strontium Titanate. *Phys. Rev.* **1964**, *134*, A981-A997.

(2) Lytle, F. W. X-Ray Diffractometry of Low-Temperature Phase Transformations in Strontium Titanate. *J. Appl. Phys*. **1964**, *35*, 2212.

(3) Ohtomo, A.; Hwang, H. A High-Mobility Electron Gas at the LaAlO$_3$/SrTiO$_3$ Heterointerface. *Nature (London)* **2004**, *427*, 423.

(4) Santander-Syro, A. F.; Copie, O.; Kondo, T.; Fortuna, F.; Pailhes, S.; Weht, R.; Qiu, X. G.; Bertran, F.; Nicolaou, A.; Taleb-Ibrahimi, A.; Le Fevre, P.; Herranz, G.; Bibes, M.; Reyren, N.; Apertet Y.; Lecoeur, P.; Barthelemy, A.; Rozenberg, M.J. Two-Dimensional Electron Gas with Universal

Subbands at the Surface of SrTiO$_3$. *Nature* **2011**, *469*, 189-193.

(5) Reyren, N.; Thiel, S.; Caviglia, A. D.; Kourkoutis, L. F.; Hammerl, G.; Richter, C.; Schneider, C. W.; Kopp, T.; Ruetschi, A. S.; Jaccard, D.; Gabay, M.; Muller, D. A.; Triscone, J.-M.; Mannhart, J. Superconducting Interfaces between Insulating Oxides. *Science* **2007**, *317*, 1196-1199.

(6) Lin, X.; Zhu, Z.; Fauqué, B.; Behnia, K. Fermi surface of the Most dilute superconductor. *Phys. Rev. X* **2013**, *3*, 021002.

(7) Lin, X.; Bridoux, G.; Gourgout, A.; Seyfarth, G.; Krämer, S.; Nardone, M.; Fauqué, B.; Behnia, K. Critical Doping for the Onset of A Two-Band Superconducting Ground State in SrTiO$_{3-\delta}$. *Phys. Rev. Lett.* **2014**, *112*, 207002.

(8) Meevasana, W.; King, P. D.; He, R. H.; Mo, S. K.; Hashimoto, M.; Tamai, A.; Songsiriritthigul, P.; Baumberger, F.; Shen, Z. X. Creation and Control of A Two-Dimensional Electron Liquid at the Bare SrTiO$_3$ Surface. *Nature materials* **2011**, *10*, 114-118.

(9) Bert, J. A.; Kalisky, B.; Bell, C.; Kim, M.; Hikita, Y.; Hwang, H. Y.; Moler, K. A. Direct Imaging of the Coexistence of Ferromagnetism and Superconductivity at the LaAlO$_3$/SrTiO$_3$ Interface. *Nature Phys.* **2011**, *7*, 767.

(10) Zhu, G. Z.; Radtke, G.; Botton, G. A. Bonding and Structure of A Reconstructed (001) Surface of SrTiO$_3$ From TEM. *Nature* **2012**, *490*, 384-387.

(11) Honig, M.; Sulpizio, J. A.; Drori, J.; Joshua, A.; Zeldov, E.; Ilani, S. Local Electrostatic Imaging of Striped Domain Order in LaAlO$_3$/SrTiO$_3$. *Nature Mater.* **2013**, *12*, 1112-1118.

(12) Cheng, G.; Tomczyk, M.; Lu, S.; Veazey, J. P.; Irvin, M. H.; Ryu, S.; Lee, H.; Eom, C.-B.; Hellberg, C. S.; Levy, J. Electron Pairing without Superconductivity. *Nature* **2015**, *521*, 196.

(13) Chen, C.; Avila, J.; Frantzeskakis, E.; Levy, A.; Asensio, M. C. Observation of A Two-Dimensional Liquid of Frohlich Polarons at the Bare SrTiO$_3$ Surface. *Nat. Commun.* **2015**, *6*, 8585.

(14) Walker, S. M.; Bruno, F. Y.; Wang, Z.; de la Torre, A.; Ricco, S.; Tamai, A.; Kim, T. K.; Hoesch, M.; Shi, M.; Bahramy, M. S.; King, P. D. C.; Baumberger, F. Carrier-Density Control of the SrTiO$_3$ (001) Surface 2D Electron Gas Studied by ARPES. *Advanced materials* **2015**, *27*, 3894-3899.

(15) Wang, Z.; McKeown Walker, S.; Tamai, A.; Wang, Y.; Ristic, Z.; Bruno, F. Y.; de la Torre, A.; Ricco, S.; Plumb, N. C.; Shi, M.; Hlawenka, P.; Sanchez-Barriga, J.; Varykhalov, A.; Kim, T. K.; Hoesch, M.; King, P. D. C.; Meevasana, W.; Diebold, U.; Mesot, J.; Moritz, B.; et al. Tailoring the Nature and Strength of Electron-Phonon Interactions in the SrTiO$_3$(001) 2D Electron Liquid. *Nature materials* **2016**, *15*, 835.

(16) Chen, Y.; Trier, F.; Kasama, T.; Christensen, D. V.; Bovet, N.; Balogh, Z. I.; Li, H.; Thyden, K. T.; Zhang, W.; Yazdi, S.; Norby, P.; Pryds, N.; Linderoth, S. Creation of High Mobility Two-Dimensional Electron Gases Via Strain Induced Polarization at An Otherwise Nonpolar Complex Oxide Interface. *Nano Lett* **2015**, *15*, 1849.

(17) Lee, P. W.; Singh, V. N.; Guo, G. Y.; Liu, H. J.; Lin, J. C.; Chu, Y. H.; Chen, C. H.; Chu, M. W. Hidden Lattice Instabilities as Origin of the Conductive Interface between Insulating LaAlO$_3$ and SrTiO$_3$. *Nat. Commun.* **2016**, *7*, 12773.

(18) Guedes, E. B.; Muff, S.; Brito, W. H.; Caputo, M.; Li, H.; Plumb, N. C.; Dil, J. H.; Radovic, M. Universal Structural Influence on the 2D Electron Gas at SrTiO$_3$ Surfaces. *Adv. Sci.* **2021**, *8*, e2100602.

(19) Geondzhian, A.; Sambri, A.; De Luca, G. M.; Di Capua, R.; Di Gennaro, E.; Betto, D.; Rossi, M.; Peng, Y. Y.; Fumagalli, R.; Brookes, N. B.; Braicovich, L.; Gilmore, K.; Ghiringhelli, G.; Salluzzo, M.

Large Polarons as Key Quasiparticles in SrTiO$_3$ and SrTiO$_3$-based heterostructures. *Phys. Rev. Lett.* **2020**, *125*, 126401.

(20) Pai, Y. Y.; Lee, H.; Lee, J. W.; Annadi, A.; Cheng, G.; Lu, S.; Tomczyk, M.; Huang, M.; Eom, C. B.; Irvin, P.; Levy, J. One-Dimensional Nature of Superconductivity at the LaAlO$_3$/SrTiO$_3$ Interface. *Phys. Rev. Lett.* **2018**, *120*, 147001.

(21) Yang, B.; Zhao, C.; Xia, B.; Ma, H.; Chen, H.; Cai, J.; HaoYang; Liu, X.; Liu, L.; Guan, D.; Wang S.; Li, Y.; Liu, C.; Zheng, H.; Jia J.. Interface-Enhanced Superconductivity in Monolayer 1T-MoTe$_2$ on SrTiO$_3$(001). *Quantum Frontiers* **2023**, *2*, 9.

(22) Christensen, D. V.; Trier, F.; Niu, W.; Gan, Y.; Zhang, Y.; Jespersen, T. S.; Chen, Y.; Pryds, N. Stimulating Oxide Heterostructures: A Review on Controlling SrTiO$_3$-Based Heterointerfaces with External Stimuli. *Adv. Mater. Interfaces* **2019**, *6*, 1900772.

(23) Erdman, N.; Poeppelmeier, K. R.; Asta, M.; Warschkow, O.; Ellis, D. E.; Marks, L. D. The Structure and Chemistry of the TiO$_2$-Rich Surface of SrTiO$_3$ (001). *Nature* **2002**, *419*, 55.

(24) Cook, S.; Marks, L. D. Ab Initio Predictions of Double-Layer TiO$_2$-Terminated SrTiO$_3$(001) Surface Reconstructions. *J. Phys. Chem. C* **2018**, *122*, 21991.

(25) Andersen, T. K.; Fong, D. D.; Marks, L. D. Pauling's Rules for Oxide Surfaces. *Surf. Sci. Rep.* **2018**, *73*, 213-232.

(26) Hamada, I.; Shimizu, R.; Ohsawa, T.; Iwaya, K.; Hashizume, T.; Tsukada, M.; Akagi, K.; Hitosugi, T. Imaging the Evolution of d States at A Strontium Titanate Surface. *J. Am. Chem. Soc.* **2014**, *136*, 17201.

(27) Song, C.; Li, X.; Jiang, Y.; Wang, X.; Yao, J.; Meng, S.; Zhang, J. Real-Space Imaging of Orbital Selectivity on SrTiO$_3$(001) Surface. *ACS Appl. Mater. Interfaces* **2019**, *11* 37279.

(28) Kienzle, D. M.; Becerra-Toledo, A. E.; Marks, L. D. Vacant-Site Octahedral Tilings on SrTiO$_3$ (001), the ($\sqrt{13}\times\sqrt{13}$)R33.7° Surface, and Related Structures. *Phys. Rev. Lett.* **2011**, *106*, 176102.

(29) Yim, C. M.; Watkins, M. B.; Wolf, M. J.; Pang, C. L.; Hermansson, K.; Thornton1, G. Engineering Polarons at A metal Oxide Surface. *Phys. Rev. Lett.* **2016**, *117*, 116402.

(30) Salman, Z.; Kiefl, R. F.; Chow, K. H.; Hossain, M. D.; Keeler, T. A.; Kreitzman, S. R.; Levy, C. D.; Miller, R. I.; Parolin, T. J.; Pearson, M. R.; Saadaoui, H.; Schultz, J. D.; Smadella, M.; Wang, D.; MacFarlane, W. A. . Near-Surface Structural Phase Transition of SrTiO$_3$ Studied with Zero-Field $\beta$-Detected Nuclear Spin Relaxation and Resonance. *Phys. Rev. Lett.* **2006**, *96*,

(31) Santander-Syro, A. F.; Fortuna, F.; Bareille, C.; Rödel, T. C.; Landolt, G.; Plumb, N. C.; Dil, H.; Radović, M. Giant Spin Splitting of the Two-Dimensional Electron Gas at the Surface of SrTiO$_3$. *Nature Mater.* **2014**, *13*, 1085.

(32) Rice, W. D.; Ambwani, P.; Bombeck, M.; Thompson, J. D.; Haugstad, G.; Leighton, C.; Crooker, S. A. Persistent Optically Induced Magnetism in Oxygen-Deficient Strontium Titanate. *Nature Mater.* **2014**, *13*, 481.

(33) Herranz, G.; Singh, G.; Bergeal, N.; Jouan, A.; Lesueur, J.; Gazquez, J.; Varela, M.; Scigaj, M.; Dix, N.; Sanchez, F.; Fontcuberta, J. Engineering Two-Dimensional Superconductivity and Rashba Spin-Orbit Coupling in LaAlO$_3$/SrTiO$_3$ Quantum Wells by Selective Orbital Occupancy. *Nat. Commun.* **2015**, *6*, 6028.

(34) Hong, C.; Zou, W.; Ran, P.; Tanaka, K.; Matzelle, M.; Chiu, W.-C.; Markiewicz, R. S.; Barbiellini, B.; Zheng, C.; Li, S.;Bansil, A.; He, R.-H. Anomalous Intense Coherent Secondary Photoemission from A Perovskite Oxide. *Nature* **2023**, *617*, 493.

(35) Lawler, M. J.; Fujita, K.; Lee, J.; Schmidt, A. R.; Kohsaka, Y.; Kim, C. K.; Eisaki, H.; Uchida, S.; Davis,

J. C.; Sethna, J. P.; Kim, E.-A. Intra-Unit-Cell Electronic Nematicity of the High-$T_c$ Copper-Oxide Pseudogap States. *Nature* **2010**, *466*, 347.

(36) Li, W.; Zhang, Y.; Deng, P.; Xu, Z.; Mo, S. K.; Yi, M.; Ding, H.; Hashimoto, M.; Moore, R. G.; Lu, D. H.; Chen, X.; Shen, Z.-X.; Xue, Q.-K.. Stripes Developed at the Strong Limit of Nematicity in FeSe Film. *Nature Physics* **2017**, *13*, 957.

(37) Hoffman, J. E.; Hudson, E. W.; Lang, K. M.; Madhavan, V.; Eisaki, H.; Uchida, S.; Davis, J. C. A Four Unit Cell Periodic Pattern of Quasi-Particle States Surrounding Vortex Cores in $Bi_2Sr_2CaCu_2O_{8+\delta}$. *Science* **2002**, *295*, 466.

(38) Hanaguri, T.; Lupien, C.; Kohsaka, Y.; Lee, D.-H.; Azuma, M.; Takano, M.; Takagi, H.; Davis, J. C. A 'Checkerboard' Electronic Crystal State in Lightly Hole-Doped $Ca_{2-x}Na_xCuO_2Cl_2$. *Nature* **2004**, *430*, 1001.

(39) Hamidian, M. H.; Edkins, S. D.; Joo, S. H.; Kostin, A.; Eisaki, H.; Uchida, S.; J. Lawler, M.; Kim, E.-A.; Mackenzie, A. P.; Fujita, K.; Lee, J.; Seamus Davis, J. C. Detection of A Cooper-Pair Density Wave in $Bi_2Sr_2CaCu_2O_{8+x}$. *Nature* **2016**, *532*, 343.

(40) Ye, S.; Zou, C.; Yan, H.; Ji, Y.; Xu, M.; Dong, Z.; Chen, Y.; Zhou, X.; Wang, Y. The Emergence of Global Phase Coherence from Local Pairing in Underdoped Cuprates. *Nature Physics* **2023**, *19*, 1301.

(41) Xue, C.-L.; Yuan, Q.-Q.; Xu, Y.-J.; Li, Q.-Y.; Dou, L.-G.; Jia, Z.-Y.; Li, S.-C. Checkerboard Ordered State in A Superconducting $FeSe/SrTiO_3(001)$ Monolayer. *Physical Review B* **2023**, *107*, 134516.

(42) Ohtomo, A.; Muller, D. A.; Grazul, J. L.; Hwang, H. Y. Artificial Charge-Modulationin Atomic-Scale Perovskite Titanate Superlattices. *Nature* **2002**, *419*, 378.

(43) C. Richter, H. B., W. Dietsche, E. Fillis-Tsirakis, R. Jany, F. Loder, L. F. Kourkoutis, D. A. Muller, J. R. Kirtley, C. W. Schneider, J. Mannhart. Interface Superconductor with Gap Behaviour Like A High-Temperature Superconductor. *Nature* **2013**, *502*, 528.

(44) Collignon, C.; Lin, X.; Rischau, C. W.; Fauqué, B.; Behnia, K. Metallicity and Superconductivity in Doped Strontium Titanate. *Annu. Rev. Condens. Matter Phys.* **2019**, *10* 25-44.

(45) Itahashi, Y. M.; Ideue, T.; Saito, Y.; Shimizu, S.; Ouchi, T.; Nojima, T.; Iwasa, Y. Nonreciprocal Transport in Gate-Induced Polar Superconductor $SrTiO_3$. *Science Advances* **2020**, *6*, eaay9120.

(46) Enderlein, C.; Oliveira, J. F. d.; Tompsett, D. A.; Saitovitch, E. B.; Saxena, S. S.; Lonzarich, G. G.; Rowley, S. E. Superconductivity Mediated by Polar Modes in Ferroelectric Metals. *Nature Commun.* **2020**, *11*, 4852.

(47) A. Stucky, G. W. S., Z. Ren, D. Jaccard, J.-M. Poumirol, C. Barreteau, E. Giannini, D. van der Marel. Isotope Effect in Superconducting n-Doped $SrTiO_3$. *Scientific Reports* **2016**, *6*, 37582.

(48) Tomioka, Y.; Shirakawa, N.; Shibuya, K.; Inoue, I. H. Enhanced Superconductivity Close to A Nonmagnetic Quantum Critical Point in Electron-Doped Strontium Titanate. *Nature Commun.* **2019**, *10*, 738.

(49) Rischau, C. W.; Pulmannová, D.; Scheerer, G. W.; Stucky, A.; Giannini, E.; Marel, D. v. d. Isotope Tuning of the Superconducting Dome of Strontium titanate. *Phys. Rev. Research* **2022**, *4*, 013019.

Supplementary Materials

**Unidirectional charge orders induced by oxygen vacancies on SrTiO$_3$(001)**

Cui Ding, Wenfeng Dong, Xiaotong Jiao, Zhiyu Zhang, Guanming Gong, Zhongxu Wei, Lili Wang, Jin-Feng Jia, and Qi-Kun Xue

Table S1 Summary of anisotropy ratios of various reconstructions at different sample biases and temperatures.

| $T/V_s$ | $\gamma=b_0/a_0$ | ($\sqrt{13} \times \sqrt{13}$) | c(4 × 2) | (2 × 1) | (2 × 2) | c(4 × 4) |
|---|---|---|---|---|---|---|
| 78 K | 0.5 V | 1-1.12 | 1.07 | 1.03-1.10 | 1-1.06 | 1-1.03 |
| 4.8 K | 0.5 V | 1-1.20 | 1.07-1.19 | 1-1.15 | 1-1.01 | ~1 |
|  | 3.5/4.0 V | 1-1.22 | 1.03-1.19 | 1-1.23 | 1-1.03 | 1-1.01 |

Table S2 Summary of densities of O$_e$ vacancies counted based on the atomically resolved images (50 nm × 50 nm) displayed in Fig. S6. The carrier density is estimated with the assumption that each oxygen vacancy denotes two electrons.

|  | ($\sqrt{13} \times \sqrt{13}$) | c(4 × 2) | (2 × 2) | c(4 × 4) | c(4 × 4)-ideal |
|---|---|---|---|---|---|
| Counts O$_v$ | 587 | 727 | 1564 | 3302 | 4110 |
| n (e$^-$/f.u.) | 0.071 | 0.088 | 0.190 | 0.402 | 0.500 |
| n (10$^{13}$ cm$^{-2}$) | 4.7 | 5.8 | 12.5 | 26.4 | 32.9 |

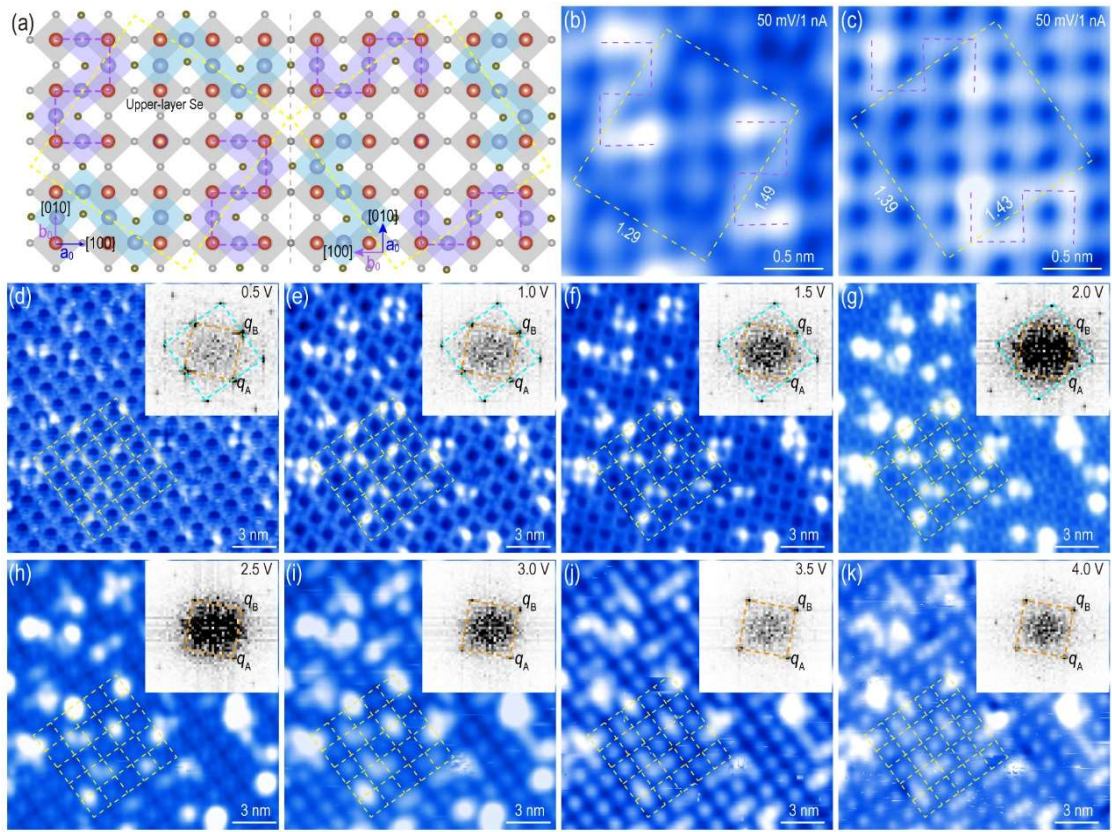

**Fig. S1** Mirror symmetry and bias-dependent modulation on (√13 × √13) surface. (a-c) Mirror symmetric unit cells and chiral TiO$_{5\square}$-pentamer. (a) Schematic of monolayer FeSe on SrTiO$_3$(001). (b) and (c) Atomically resolved Se(001) lattices with mirrored "2" and "5" like DOS contrasting patterns, respectively. The chirality reversal in TiO$_{5\square}$ pentamer is likely due to the inherent anti-clockwise rotations of adjacent TiO$_6$ octahedrons under antiferrodistortive transition. (d-k) Bias-dependent atomically resolved images with respective FFT images inserted showing the (√2$q_A$/2, √2$q_B$/2) modulation under $V_s$ < 2.0 V.

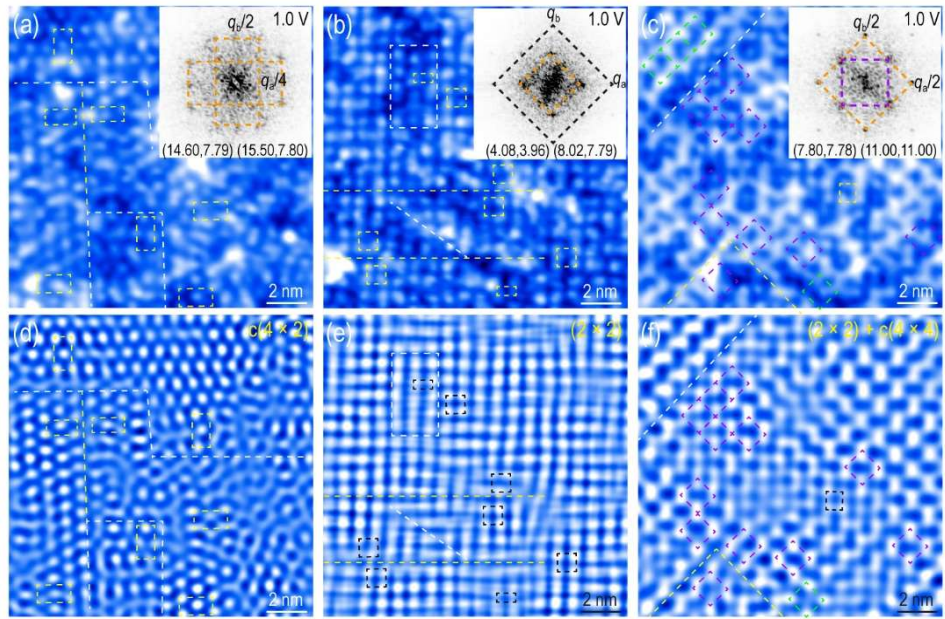

**Fig. S2** LN-temperature STM characterization of SrTiO$_3$(001)- c(4 × 2), -(2 × 2) and -(2 × 2) + c(4 × 4) surfaces. (a, b, c) Atomically resolved topographic images taken on terraces with the respective FFT images inserted, and (d, e, f) the corresponding inverted FFT images generated from all the Bragg spots. The numbers at the bottom of FFT images are the deduced ($2n\pi/q_a$, $2m\pi/q_b$) values in the units of Å. The yellow/black dashed squares/rectangles mark the respective unit cells, the white (yellow) dashed lines mark the twin (anti-phase) boundaries, and the purple and green dashed diamonds mark the c(4 × 4) orders.

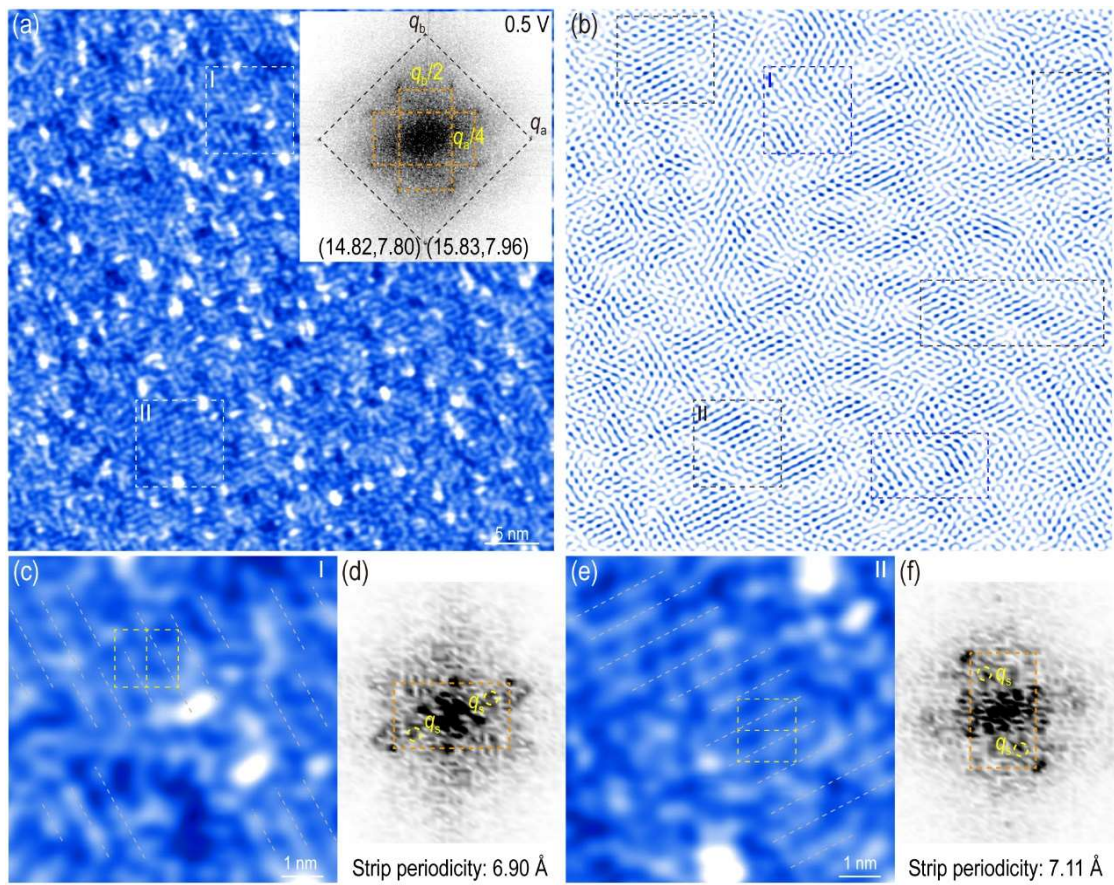

**Fig. S3** Unidirecitional stripe order on the SrTiO$_3$(001)- c(4 × 2) surface at LHe-temperature and under small sample bias $V_s$ = 0.5 V. (a) Large-scale atomically resolved topographic image with the FFT image inserted. (b) The inverted-FFT image generated from the two sets of Bragg spots. (c, e) Zoom-in atomically resolved topographic images of the I and II regions marked in (a), and (d,f) the corresponding FFT patterns, with the yellow dashed circles marking the Bragg spots corresponding to the stripe orders.

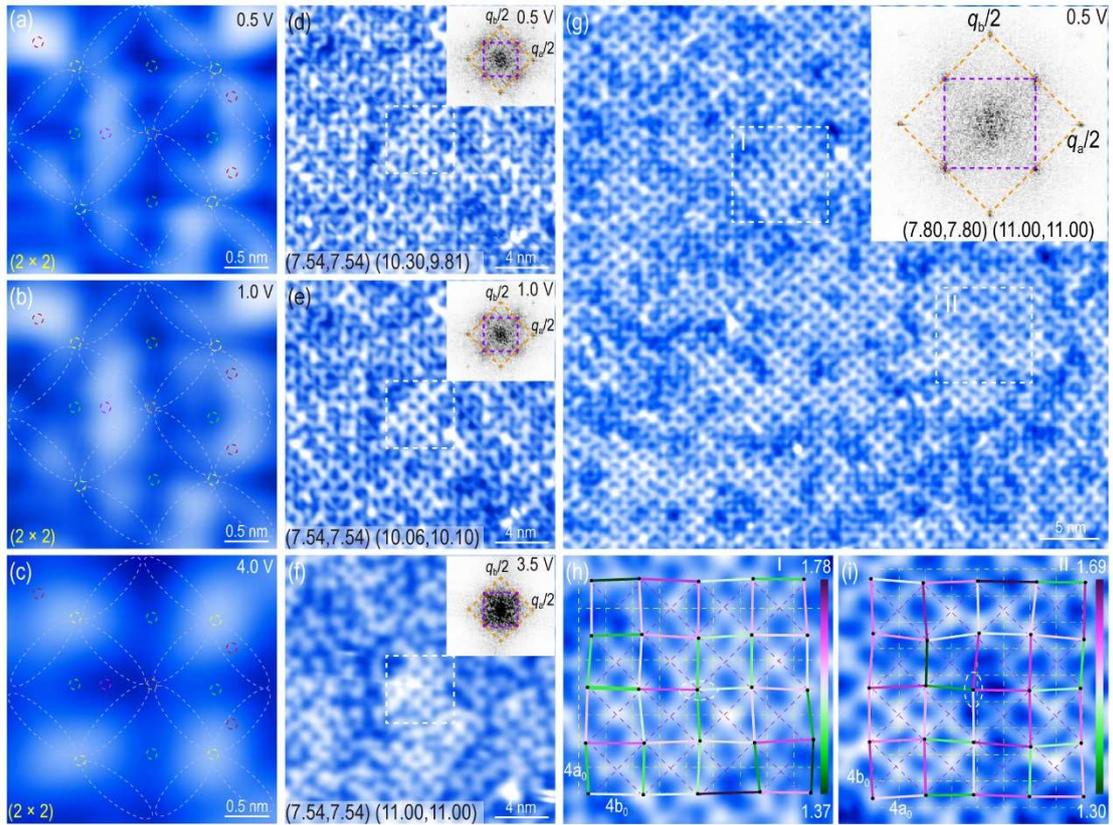

**Fig. S4** LHe-temperature STM characterization of SrTiO$_3$(001)-(2 × 2) surfaces. (a,b,c) Bias-dependent atomically resolved images of (2 × 2) surface without c(4 × 4). (d, e, f) Bias-dependent atomically resolved images of (2 × 2) surface with c (4 × 4) order. The white dashed squares mark the regions that the zoom-in images displayed in Figs. 4(d,e,f). (g) Large-scale topographic image showing domains of c (4 × 4) order with revered a$_0$/b$_0$ in I and II regions. (h) and (i) the zoom-in images in I and II regions consistently show bound TiO$_{5\square}$ (exemplified by purple ovals) along the large lattice direction b$_0$. The statistics (4a$_0$, 4b$_0$) values for Regions I and II are (15.30 ± 0.76, 15.85 ± 1.00) and (15.20 ± 0.93, 15.47 ± 0.88), respectively, and correspondingly, the anisotropy ratio γ=b$_0$/a$_0$ is 1.04 and 1.02.

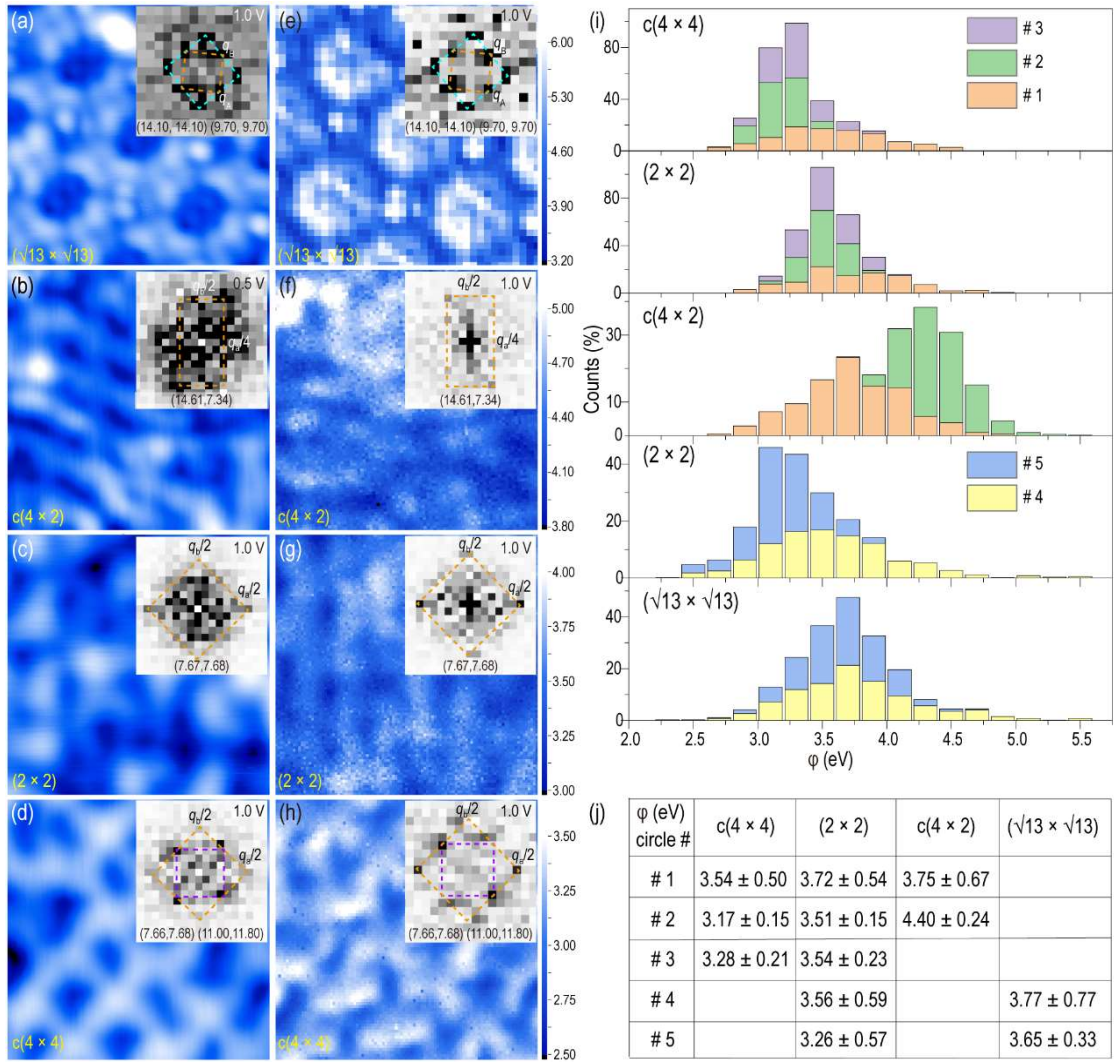

**Fig. S5** LHe-temperature work function characterizations of SrTiO$_3$(001)- (√13 × √13), c(4 × 2), (2 × 2) and c(4 × 4) surfaces. (a-d) topographic images and (e-h) the corresponding work function mapping images of the same regions, with the respective FFT patterns inserted. (i) Statistical comparison of work function values of various reconstructed surfaces obtained by the same tip with different circles of tip treatments as labeled, and (j) summary of the average work function values.

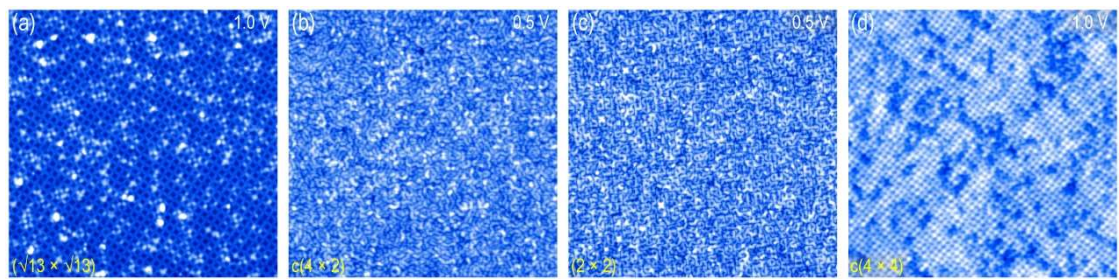

**Fig. S6** LHe-temperature atomically resolved images on various surfaces for the calculations of oxygen vacancy densities summarized in Table S2.

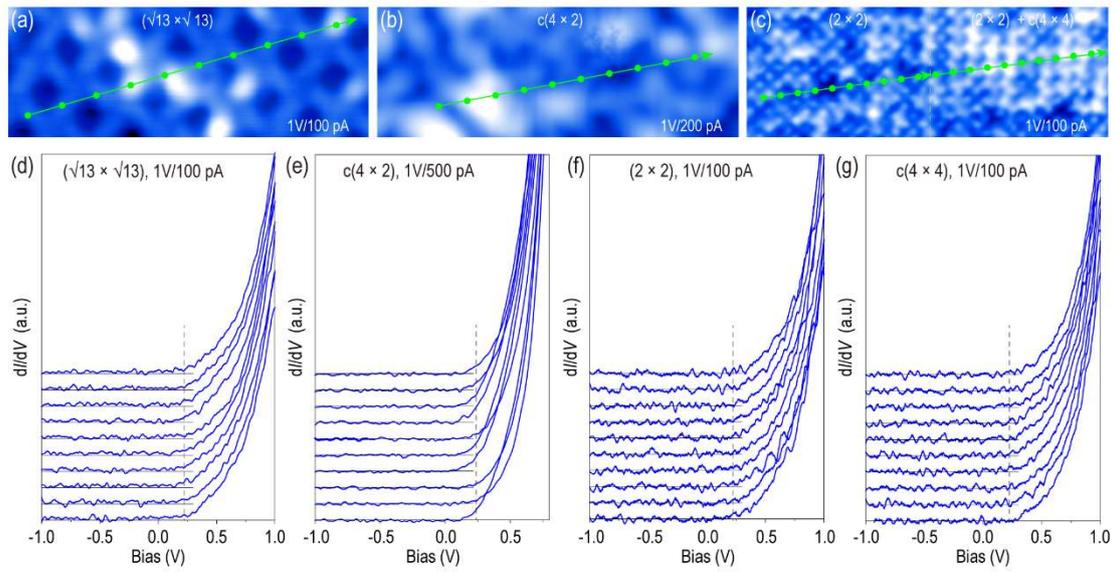

**Fig. S7** LHe-temperature d$I$/d$V$ tunneling spectra of SrTiO$_3$(001)-surfaces with various reconstructions. (a,b,c) Topographic images of SrTiO$_3$(001) -(√13 × √13), c (4 × 2) and (2 × 2) + c (4 × 4) surfaces. (d,e,f,g) d$I$/d$V$ tunneling spectra taken on the various reconstructions at the points marked in (a-c).